\documentclass{moriond}

\usepackage{ifthen}
\newboolean{enable-backrefs} 
\setboolean{enable-backrefs}{false} 
\newboolean{enable-lineno} 
\setboolean{enable-lineno}{false} 
\newboolean{enable-printout} 
\setboolean{enable-printout}{true} 

\PassOptionsToPackage{ngerman,main=UKenglish}{babel}   
\usepackage{babel}					
\usepackage[super]{nth} 

\PassOptionsToPackage{pagewise,displaymath}{lineno} 
\usepackage{lineno} 
\modulolinenumbers[1] 
\ifthenelse{\boolean{enable-lineno}}%
{%
   \newcommand*\patchAmsMathEnvironmentForLineno[1]{%
	  \expandafter\let\csname old#1\expandafter\endcsname\csname #1\endcsname
	  \expandafter\let\csname oldend#1\expandafter\endcsname\csname end#1\endcsname
	  \renewenvironment{#1}%
	  {\linenomath\csname old#1\endcsname}%
   {\csname oldend#1\endcsname\endlinenomath}}%
   \newcommand*\patchBothAmsMathEnvironmentsForLineno[1]{%
	  \patchAmsMathEnvironmentForLineno{#1}%
   \patchAmsMathEnvironmentForLineno{#1*}}%
   \AtBeginDocument{%
	  \patchBothAmsMathEnvironmentsForLineno{equation}%
	  \patchBothAmsMathEnvironmentsForLineno{align}%
	  \patchBothAmsMathEnvironmentsForLineno{flalign}%
	  \patchBothAmsMathEnvironmentsForLineno{alignat}%
	  \patchBothAmsMathEnvironmentsForLineno{gather}%
	  \patchBothAmsMathEnvironmentsForLineno{multline}%
   }
   \linenumbers 
}{\relax}

\ifthenelse{\boolean{enable-backrefs}}%
{%
   \ExecuteBibliographyOptions{backref=true}
}{\relax}

\usepackage{amsmath}
\usepackage{bbm}
\usepackage{amsthm}
\usepackage{mathtools}

\usepackage{xfrac} 
\usepackage{nicefrac} 
\usepackage[makeroom]{cancel}
\usepackage{xparse} 
\usepackage{physics} 

\usepackage{siunitx} 
\usepackage{suffix} 
\usepackage{siunitx_extension} 

\usepackage{booktabs} 
\usepackage{adjustbox} 
\usepackage{multicol}
\usepackage{multirow}
\usepackage{tabulary}
\usepackage{tabularx} 
\usepackage{enumitem} 
\usepackage{subfig}  
\usepackage{floatpag}
\usepackage{caption}
\usepackage{wrapfig}
\usepackage{rotating}

\usepackage{scrhack} 
\usepackage{mparhack} 

\PassOptionsToPackage{novbox}{pdfsync} 
\usepackage{pdfsync} 
\PassOptionsToPackage{
   pdftex,
   hyperfootnotes=false,
   pdfpagelabels
}{hyperref}
\usepackage{hyperref}
\hypersetup{%
   pdfstartpage=3, pdfstartview=FitV,%
   colorlinks=true, linktocpage=true, 
   breaklinks=true, pdfpagemode=UseNone, pageanchor=true, pdfpagemode=UseOutlines,%
   plainpages=false, bookmarksnumbered, bookmarksopen=true, bookmarksopenlevel=1,%
   hypertexnames=true, pdfhighlight=/O,
   urlcolor=webbrown, linkcolor=RoyalBlue, citecolor=webgreen, 
   pdfsubject={},%
   pdfkeywords={},%
   pdfcreator={pdfLaTeX},%
   pdfproducer={LaTeX with hyperref and classicthesis}%
}   
\ifthenelse{\boolean{enable-printout}}%
{%
   \hypersetup{%
	  colorlinks=false, linktocpage=false, pdfborder={0 0 0}, pdfstartpage=3, pdfstartview=FitV,%
   }
}{\relax}

\PassOptionsToPackage{pdftex}{graphicx}
\usepackage{graphicx} 
\DeclareGraphicsExtensions{.pdf,.png,.gif,.jpg,.jpeg} 
\graphicspath{{figures/}}

\PassOptionsToPackage{english}{cleveref}   
\usepackage{cleveref} 
\Crefname{figure}{figure}{Figure}

\PassOptionsToPackage{usenames,dvipsnames,svgnames,table}{xcolor} 
\usepackage{xcolor}
\definecolor{webgreen}{rgb}{0,.5,0} 
\definecolor{webbrown}{rgb}{.6,0,0} 

\usepackage{xspace} 
\usepackage{url} 
\usepackage{blindtext} 
\usepackage{umoline} 

\usepackage{tikz}
\usetikzlibrary{calc,
   tikzmark,
   plotmarks,
   arrows,
   arrows.meta,
   shapes,
   decorations,
   automata,
   backgrounds,
   petri,
   trees,
   patterns,
   fit,
   positioning,
   fadings,
   decorations.pathmorphing,
   decorations.markings,
}
\usepackage{environ}
\makeatletter
\newsavebox{\measure@tikzpicture}
\NewEnviron{scaletikzpicturetowidth}[1]{
   \def\tikz@width{#1}%
   \begin{lrbox}{\measure@tikzpicture}%
	  \BODY
   \end{lrbox}%
   \pgfmathparse{#1/\wd\measure@tikzpicture}%
   \BODY
}
\tikzset{cross/.style={cross out, draw, 
	  minimum size=2*(#1-\pgflinewidth), 
inner sep=0pt, outer sep=0pt}}
\definecolor{turquoise}{rgb}{0 206 209}
\makeatother
\usepackage{pgfplots}
\pgfplotsset{compat=newest} 
\usepackage{pgfplotstable}

\PassOptionsToPackage{
   colorinlistoftodos,
   textwidth=2.5cm,
   textsize=scriptsize
}{todonotes}
\usepackage{todonotes} 

\input{alias}
\def\be{\begin{equation}}
\def\ee{\end{equation}}
\def\bea{\begin{eqnarray}}
\def\eea{\end{eqnarray}}

\newcommand{\Photo}{}

\begin{document}
\vspace*{4cm}
\title{Charm physics at \bes}

\author{ P. Weidenkaff \footnote{weidenka@uni-mainz.de}\\ on behalf of the \bes collaboration}

\address{Institut f\"ur Kernphysik, Johann-Joachim-Becher-Weg 45,\\
55128 Mainz, Germany}

\maketitle\abstracts{
   The study of mesons and baryons which contain at least one charm quark is referred to as open charm physics. It offers the possibility to study up-type quark transitions. Since the \c quark can not be treated in any mass limit, theoretical predictions are difficult and experimental input is crucial.
   \bes collected large data samples of \epem collisions at several charm thresholds. The at-threshold decay topology offers special opportunities to study open charm decays.
   We present a selection of recent \bes results. Branching fractions and the \Ds decay constant are measured using the leptonic decays to $\mup\nu$ and $\taup\nu$. From a data sample of \SI{0.482}{\invfb} collected at the \DsDsb threshold we measure $f_{\Ds}=\SIerrs{241}{16.3}{6.6}{\MeV}$. \bes recently found preliminary evidence of the decay $\Dp\to\taup\nut$ and with a significance larger than \SI{4}{\stdDev} using \SI{2.81}{\invfb} of data at the \DzDzb threshold.
   Using the sample data sample the decay $\Dz\to\KSL\piz(\piz)$ is analysed. The branching fractions are measured and using the \CP eigenstates $\KSL\piz$ the \Dz mixing parameter $y_{\CP} = \SI{0.98(243)}{\percent}$ is measured.
   }

\section{Introduction}
The \bes experiment is located at the Institute of High Energy Physics in Beijing. Symmetric \epem collisions from Beijing Electron-Positron Collider (BEPCII) in an energy range between \SI{2.0}{\GeV} and \SI{4.6}{\GeV} are analyzed. The maximum luminosity of BEPCII of \SI{1e33}{\Lumi} at $\sqrt{s}=$\SI{3.773}{\GeV} were surpassed in April 2016. 

The detector measures charged track momenta with a relative precision of \SI{0.5}{\percent} (@\SI{1.0}{\GeV/\c}) using a multi-wire drift chamber in a \SI{1}{\tesla} magnetic field. Electromagnetic showers are measured in a caesium iodide calorimeter with a relative precision of \SI{2.5}{\percent} (@\SI{1.0}{\GeV}) and a good particle identification is achieved by combining information from energy loss in the drift chamber, from the time-of-flight system and from the calorimeter. Muons can be identified using 9 layers of resistive plate chambers integrated in the magnet return yoke. Details are provided elsewhere \cite{Ablikim:2009aa}.
\bes has collected large data samples in the tau-charm region. The interesting samples for the study of charmed hadrons are usually at a center-of-mass energy close to a threshold. The samples of interest for the analyses described in the following were recorded at the \DzDzb/\DpDm threshold $(\sqrt{s}=\SI{3.773}{\GeV})$ and at the \DsDsb threshold $(\sqrt{s}=\SI{4.009}{\GeV})$. Integrated luminosities of \SI{2.81}{\invfb} and \SI{0.482}{\invfb} were recorded, respectively.

\begin{figure}[tbp]
   \centering
   \begin{tikzpicture}
	  \draw[draw=none, use as bounding box](0,1.5) rectangle (10.2cm,6cm);
	  \begin{scope}[scale=0.5,color=black!70!white]
		 \coordinate (origin) at (9,9);
		 \draw[->] (origin) -- +(0:1) node[above] (n1) {z};
		 \draw[->] (origin) -- +(90:1) node[left] (n2) {y};
	  \end{scope}
	  \begin{scope}[line width=2pt]
		 \node[draw,black,rectangle,rounded corners=3pt,minimum size=3pt] (pv) at (5,3) {\psiprpr};
		 \draw[blue,->]  (1,3) node[below] {\ep} -- (pv);
		 \draw[blue,->]  (9,3) node[above] {\en} -- (pv);
		 \node[draw,red,circle,minimum size=3pt, label=below:$\Dzb_{tag}$] (Dbarvtx) at (6.5,2.5) {};
		 \draw[black,dashed,-]  (pv) -- (Dbarvtx);
		 \draw[->,line width=1.5pt,black!60!white]  (Dbarvtx) -- +(5:2cm); 
		 \draw[->,line width=1.5pt,black!60!white]  (Dbarvtx) -- +(-5:2cm) node[right] {hadrons}; 
		 \draw[->,line width=1.5pt,black!60!white]  (Dbarvtx) -- +(-15:2cm); 
		 \draw[->,line width=1.5pt,black!60!white]  (Dbarvtx) -- +(-25:2cm); 
		 \node[draw,red,circle,minimum size=3pt, label=above right:$\Dz$] (Dvtx) at (3.5,3.5) {};
		 \node[draw,black!60!white,fill,circle,minimum size=0pt] (KSvtx) at ($ (Dvtx) + (182:1.5) $) {};
		 \draw[black!60!white,dashed,-]  (Dvtx) -- (KSvtx);
		 \draw[black,dashed,-]  (pv) -- (Dvtx);
		 \draw[black!60!white]  (KSvtx) -- +(190:1) node[left] {$l^+$};
		 \draw[black!60!white,dashed]  (KSvtx) -- +(170:1) node[left] {$\nu_l$};
		 \draw[black!60!white]  (Dvtx) -- +(150:2) node[left] {};
		 \draw[black!60!white,-]  (Dvtx) -- +(120:2) node[left] {hadrons};
	  \end{scope}
   \end{tikzpicture}
   \caption{\psiprpr decay topology in the \psiprpr rest frame. An undetected particle track can be reconstructed using the constrained kinematics of the decay. Typical tag modes for \CP and flavour eigenstates are listed.}
   \label{fig:psiprprdecay}
\end{figure}
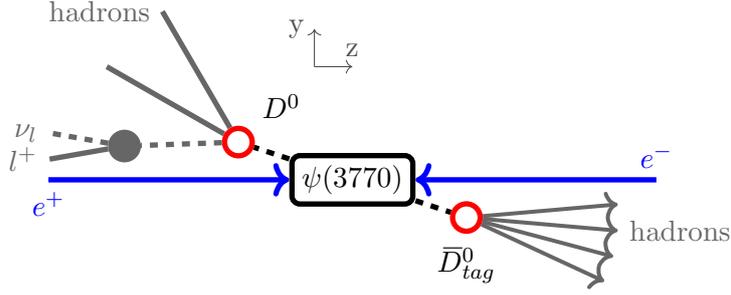
The at-threshold decay topology at a center-of-mass energy of \SI{3.773}{\GeV} is illustrated in \cref{fig:psiprprdecay}. A pair of mesons is produced and it is possible to conclude from the decay of one meson (so-called tag meson) properties of the second decay. For instance in case of neutral \D decays the flavour or the \CP quantum numbers of the signal decay can be measured, even if the signal final state does not provide this information. In case of charged \D decays the reconstruction of both decays is used to reduce the background and furthermore if undetected particles are involved in the signal decay the four momenta of those can be reconstructed. In particular the study of leptonic and semi-leptonic decays benefits from this. The reconstruction of both decays in each event is referred to as double tag technique.

In the following we present the measurements of the \Ds decay constant (\cref{sec:dsmunu}), first evidence of the decay $\Dp\to\tau\nut$ (\cref{sec:dptaunu}) and the analysis of the decay $\Dz\to\KSL\piz(\piz)$ (\cref{sec:dzkspizpiz}).
\section{Pure leptonic $\Dsd^+$ decays}
\label{sec:pureLeptonicD}
The pure leptonic decay of charged \Dsd mesons proceeds via the annihilation of \c and \sbar(\dbar) to a virtual \Wpm boson and its decay to $l^+\nu_l$. The decay rate can be parametrized as:
\begin{align}
	\Gamma(\Dsd\to l^+\nu_l) = \frac{G_F^2}{8\pi} f_{\Dsd}^2 m_l^2 m_{\Dsd} \left( 1-\frac{m_l^2}{m^2_{\Dsd}}\right)^2 \abs{V_{cs(d)}}^2.
   \label{eqn:ds:decayRate}
\end{align}
With the Fermi constant $G_F$, the lepton mass $m_l$, the corresponding CKM matrix element $\abs{V_{cs(d)}}^2$, the \Dsd mass $m_{\Dsd}$ and the decay constant $f_{\Dsd}^2$. The decay constant parametrizes the \qcd effects on the decay. From the measurement of the decay width $\Gamma(\Dsd\to l^+\nu_l)$ the decay constant $f_{\Dsd}^2$ can be extracted. 

The branching fraction can be measured via the previously described double tag technique. In each event the tag decay is reconstructed via numerous decay channels. The number of events that contain a tag candidate is denoted by $N_{\text{tag}}$. Among those events the signal decay is reconstructed and the number of events that contain a tag decay and a signal decay is denoted by $N_{\text{sig,tag}}$. The branching fraction is given by:
\begin{align}
   \BR(\Dsd\to l^+\nu_l) = \frac{N_{\text{sig,tag}}}{\epsilon_{\text{sig,tag}}}\times\frac{\epsilon_{\text{tag}}}{N_{\text{tag}}}.
   \label{eqn:ds:bf}
\end{align}
The efficiencies for reconstruction and selection $\epsilon_i$ are obtained from simulation.
Since the final state contains a neutrino which is not detected the signal yield is determined using the missing mass:
\begin{align}
   MM^2 = \frac{\left(E_{\text{beam}}-E_\mu\right)^2}{c^4} - \frac{\left(-\vec{p}_{\Dsd}-\vec{p}_{\mup}\right)^2}{\c^2}.
\end{align}
The beam energy is denoted by $E_{\text{beam}}$ and the reconstructed momentum of the tag \Dsd decay candidate by $\vec{p}_{\Dsd}$.

\pagebreak
\subsection{$\Ds\to\mu^+\numu$ and $\Ds\to\tau^+\nut$}
\label{sec:dsmunu}
\begin{wrapfigure}[21]{r}{0.35\textwidth}
   \centering
   \includegraphics[width=0.35\textwidth]{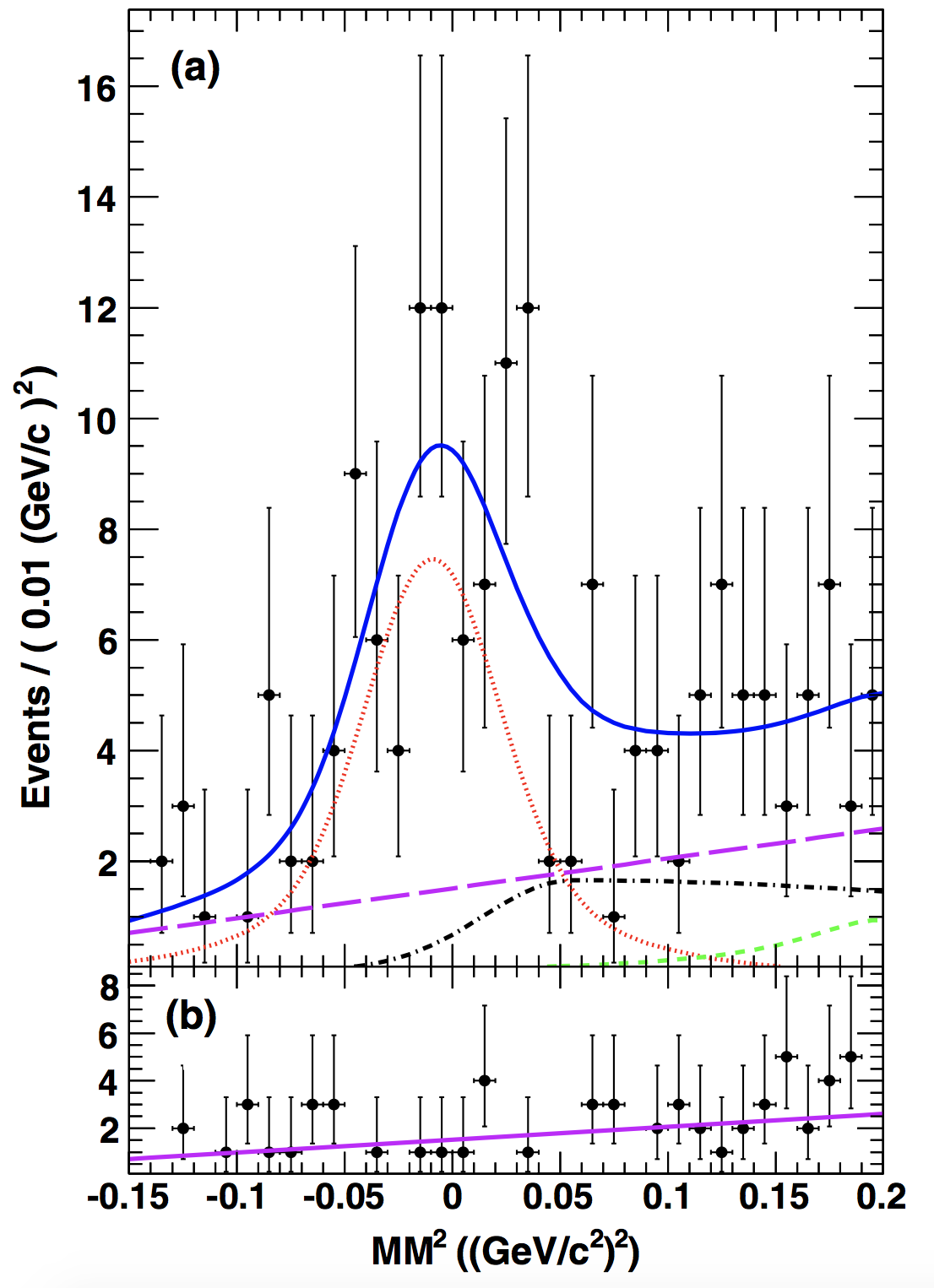}
   \caption{$MM^2$ distribution of $\Ds\to\mu^+\numu$ and $\Ds\to\tau^+\nut$. Signal (a) and sideband (b) regions are shown.}
   \label{fig:ds:missingMass}
\end{wrapfigure}

The distribution of $MM^2$ of $\Ds\to\mu^+\numu$ and $\Ds\to\tau^+\nut$ is shown in \cref{fig:ds:missingMass}. The \taup is reconstructed via its decay to $\pip\nutb$. The yield is determined via a simultaneous fit to signal and sideband regions whereas the sideband regions are defined in the \Dsb mass spectrum of the tag candidate. The $\mup\numu$ signal is shown as red dotted curve and the $\taup\nut$ signal as black dot-dashed curve. Background from misreconstructed tag \Ds decays and background from non-\DsDsb events is shown as green short dashed and violet long dashed curve, respectively.
Within a sample of \num{15127(321)} events which contain a tag candidate we find \num{69.3(93)} $\Ds\to\mup\numu$ decays and \num{32.5(43)} $\Ds\to\taup\nut$ decays. In the fitting procedure the ratio of $\Ds\to\mup\numu$ to $\Ds\to\taup\nut$ was constraint to its Standard model prediction. The yields are corrected for radiative effects and we obtain:
\begin{align}
   \BR(\Ds\to\mup\numu) &= \SIerrs{0.495}{0.067}{0.026}{\percent} \nonumber\\
   \BR(\Ds\to\taup\nut) &= \SIerrs{4.83}{0.65}{0.26}{\percent}.
\end{align}
The branching fractions  \BR($\Ds\to\mup\numu$) and  \BR($\Ds\to\taup\nut$) are consistent with the world average within \num{1} and \num{1.5} standard deviations, respectively.
Furthermore, the branching fractions are consistently determined using a fitting method which does not rely on the ratio of $\Ds\to\mup\numu$ to $\Ds\to\taup\nut$. For further details we refer to \cite{Ablikim:2016duz}.

Using $\BR(\Ds\to\mup\numu)$ the decay constant $f_{\Ds}$ is determined using \cref{eqn:ds:decayRate}:
\begin{align}
   f_{\Ds} = \SIerrs{241.0}{16.3}{6.5}{\MeV}.
\end{align}
The CKM matrix element $\abs{V_{cd}}=\num{0.97425(22)}$ \cite{Agashe:2014kda} and the \Ds lifetime \cite{Agashe:2014kda} is used. A good agreement with LQCD calculations is found. Result are published in \cite{Ablikim:2016duz}

\subsection{$\Dp\to\tau^+\nut$}
\label{sec:dptaunu}
\begin{figure}[tb]
	\centering
	\subfloat[$E_{EMC} \leq \SI{300}{\MeV}$]{
	\includegraphics[width=0.45\textwidth]{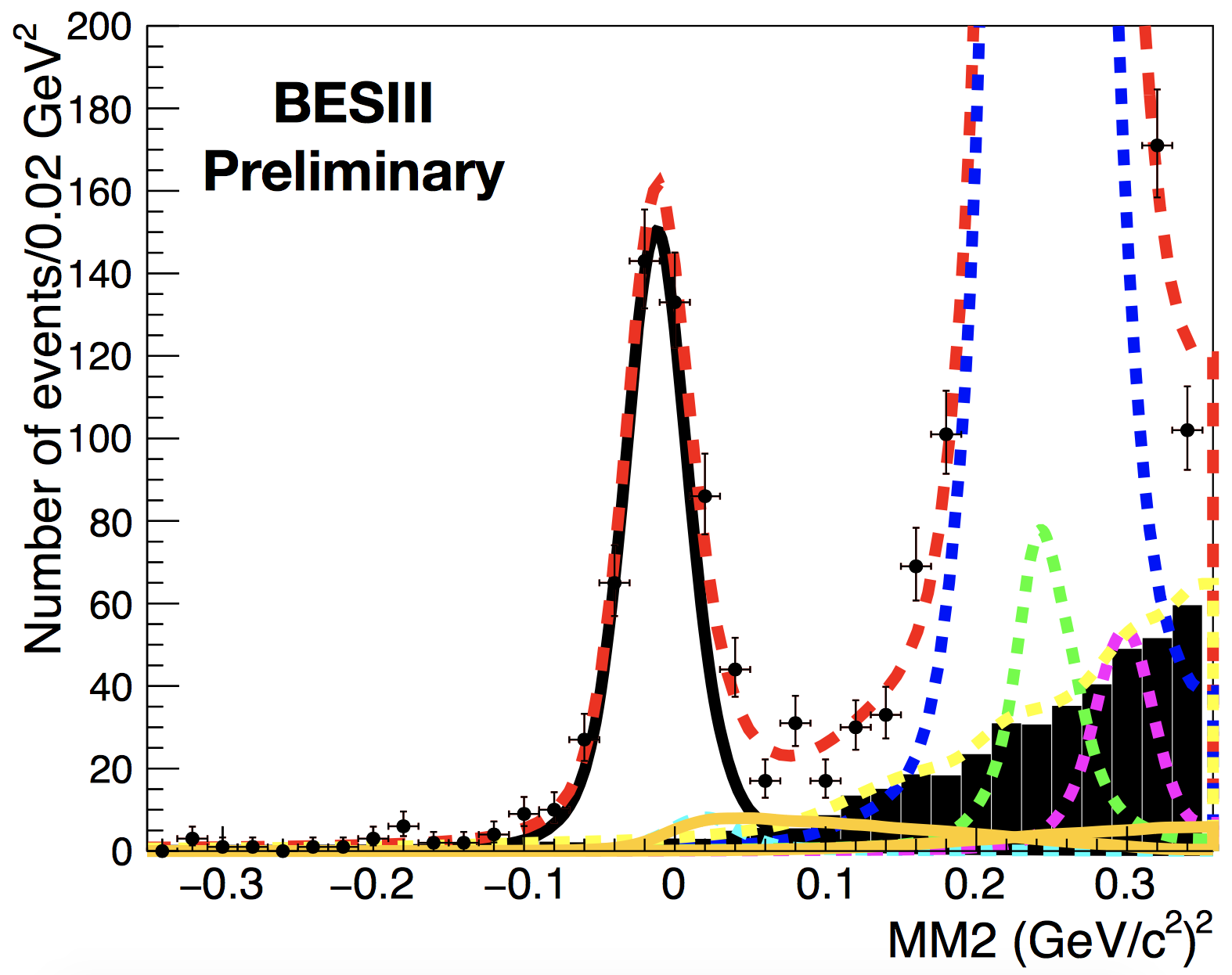}}
	\subfloat[$E_{EMC} > \SI{300}{\MeV}$]{
	\includegraphics[width=0.45\textwidth]{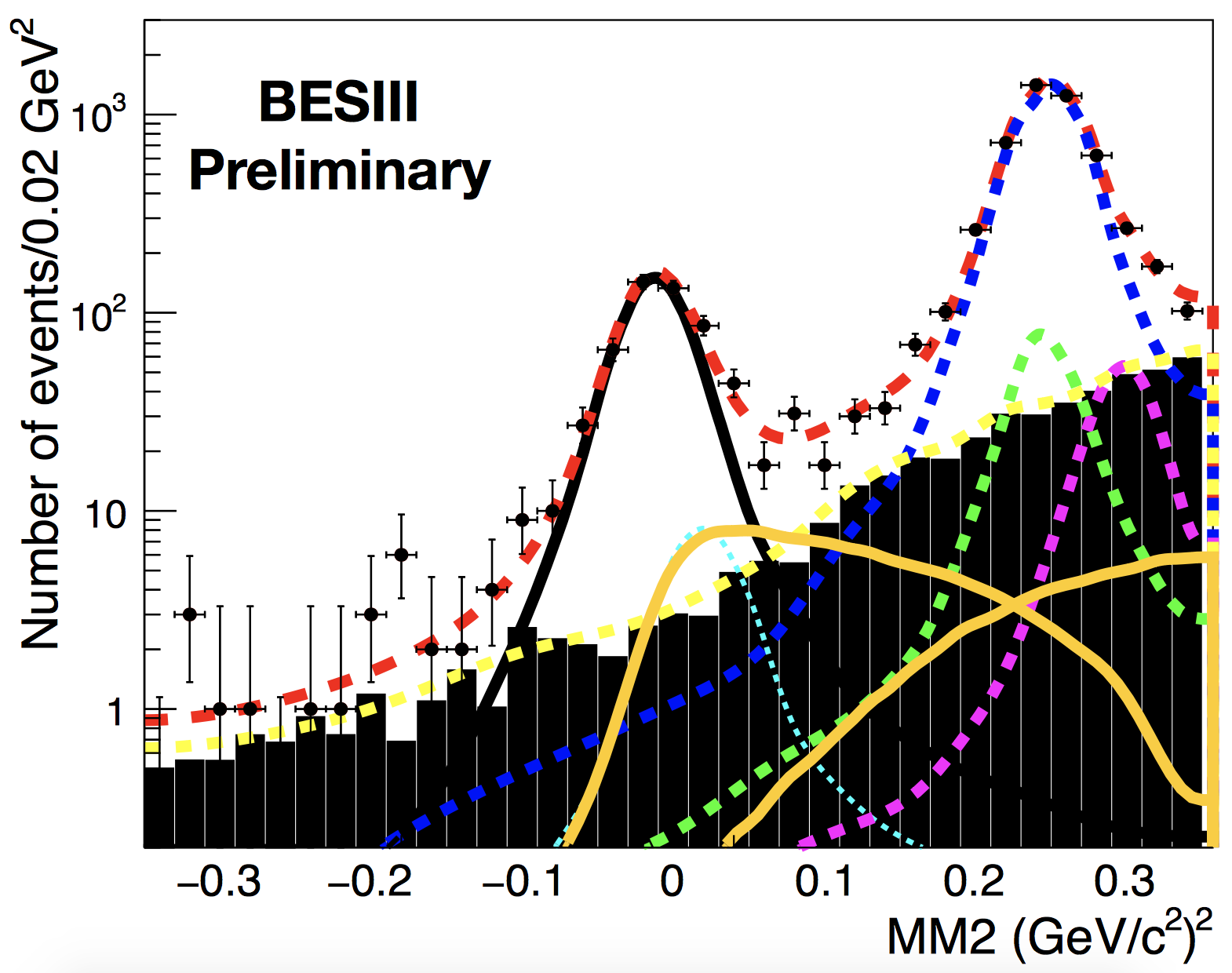}}
	\caption{$MM^2$ distribution for the decay $\Dp\to\tau^+\nut$. The signal is shown as solid orange line. Background comes mainly from \Dp decays to $\mu^+\numu$ (solid black) and to $\pip\KL$ (dashed blue).}
	\label{fig:MMsqDptaunu}
\end{figure}

The $MM^2$ distribution of $\Dp\to\tau^+\nut$ is shown in \cref{fig:MMsqDptaunu}. The most severe background to the signal channel is $\mu^+\numu$. To distinguish signal and background in a fitting procedure we use the difference in energy deposit of pions and muons in the electromagnetic calorimeter (EMC). We split the sample into events with an energy deposit larger and sample $\SI{300}{\MeV}$. As shown in \cref{fig:MMsqDptaunu}(b) above $\SI{300}{\MeV}$ the number of $\mu^+\numu$ events is reduced compared to the number of $\tau^+\nut$ events.

We obtain a preliminary signal yield of \SI{137(27)} events. The significance of the signal is larger than \SI{4}{\stdDev}. The preliminary branching fraction is given by:
\begin{align}
	\BR(\Dp\to\tau^+\nut) = \SI{1.20(24)}{\timesten{-3}}.
\end{align}
Furthermore, we extract the ratio of $\taup\nut$ to $\mup\numu$ decays:
\begin{align}
	R := \frac{\Gamma(\Dp\to\taup\nut)}{\Gamma(\Dp\to\mup\numu)} = \num{3.21(64)}. 
\end{align}
The result is consistent with the Standard Model prediction.

\section{Analysis of the decay $\Dz\to\KSL\piz(\piz)$}
\label{sec:dzkspizpiz}
We present preliminary results of the branching measurement of the decays $\Dz\to\KSL\piz$ and $\Dz\to\KSL\piz$. Furthermore we determine the \Dz mixing parameter $y_{\CP}$ using the \CP eigenstates $\KS\piz$ and $\KL\piz$. The challenge in this channel is the reconstruction of the \KL decay since its long decay time signals of its decay products in the drift chamber is very unlikely. We use the constraint kinematics at the \DzDzb threshold to predict the \KL four-momentum and furthermore require a certain energy deposit in the electromagnetic calorimeter.

The branching fraction of a \CP eigenstate can be measured in a self-normalization way using Cabibbo favoured (CF) tag channels. We define:
\begin{align}
	M^\pm = \frac{N_{CF,CP\pm}}{\epsilon_{CF,CP\pm}}\frac{\epsilon_{CF}}{N_{CF}}.	
\end{align}
The yields of double of double and single tag events is denoted by $N_{CF,CP\pm}$ and $N_{CF}$ and the corresponding reconstruction efficiencies by $\epsilon_{CF,CP\pm}$ and $\epsilon_{CF}$.
The branching fraction is given by:
\begin{align}
	\BR_{\CP\pm} = \frac{1}{1\mp C_{f}} M^\pm, \qquad C_f = \frac{M^- - M^+}{M^- + M^+}.
\end{align}

We use the flavour tag channels $\Km\pip$, $\Km\pip\pim\pip$ and $\Km\pip\piz$. The double tag yields and the preliminary branching fractions are listed in \cref{tab:dzkspizpizBF}. The branching fractions of the final states $\KSL\piz$ and $\KS\piz\piz$ are consistent with the PDG average \cite{Agashe:2014kda} and the branching fraction to $\KS\piz\piz$ is the first accurate measurement.

From the branching fractions we can calculate the asymmetry between the \CP eigenstates:
\begin{align}
	R_{\Kz\piz(\piz)} = \frac{\BR_{\KS\piz(\piz)}-\BR_{\KL\piz(\piz)}}{\BR_{\KS\piz(\piz)}+\BR_{\KL\piz(\piz)}}.
\end{align}
The results are also listed in \cref{tab:dzkspizpizBF}.
\begin{table}[tbp]
	\centering
	\caption{Double tag yields and branching fractions of \CP eigenstates $\KSL\piz(\piz)$. Uncertainties are statistical only.}
	\label{tab:dzkspizpizBF}
	\begin{tabular}{|c|c|c|c|c|}
		\hline
		Channel & \CP &$N_{CF,CP\pm}$&  $\BR_{\CP\pm}$& R\\
		\hline
		\KS\piz & $+$ &\num{7141(91)}& \num{1.230(020)} & \multirow{2}{*}{\num{0.1077(125)}} \\
		\KL\piz & $-$ &\num{6678(118)}& \num{0.991(019)} &  \\
		\hline
		\KS\piz\piz & $+$ &\num{2623(60)}& \num{0.975(24)} & \multirow{2}{*}{\num{-0.0929(209)}} \\
		\KL\piz\piz & $-$ &\num{2136(69)}& \num{1.18(04)} &  \\
		\hline
	\end{tabular}
\end{table}

\pagebreak
\subsection{Measurement of $y_{CP}$}
Using the final states $\KS\piz$ and $\KL\piz$ we determine the \Dz mixing parameter $y_{\CP}$. The branching ratio of a \CP eigenstate is connected to the branching ratio of a pure flavour eigenstate via:
\begin{align}
	\BR_{\CP} \approx \BR_{\text{flavour}} (1\mp y_{\CP}). 	
\end{align}
The parameter $y_{\CP}$ is then given by the asymmetry of branching ratios of \CP even and odd states to pure flavour states f:
\begin{align}
	y_{\CP} = \frac{\BR_{-;f} - \BR_{+;f}}{\BR_{-;f} + \BR_{+;f}}.
\end{align}
The previously mentioned Cabibbo favoured final states are not pure flavour eigenstates. Therefore, we use the semi-leptonic decay to $\Km e^+\nue$.
We obtain a preliminary value of:
\begin{align}
	y_{\CP} = \SI{0.98(243)}{\percent}.
\end{align}
We quote statistical uncertainty only. The result is in agreement with a previous measurement of \bes \cite{Ablikim:2015hih} as well as with the HFAG average \cite{arXiv:1612.07233}. Results are preliminary and we quote statistical uncertainties only.

\section{Summary}
The \bes experiment has collected large data sample at charm-related thresholds. The constraint kinematics at those energies allow the reconstruction of (semi-) leptonic decays with low background. Furthermore, the quantum entanglement of \DzDzb at threshold provides a unique laboratory for the analysis of \CP eigenstates. 
We present the analysis of the leptonic decay of \Ds to $\mup\numu$ and $\taup\nut$ with the measurement of branching fractions of the derived \Ds form factor. Recently, \bes has found preliminary evidence of the decay \Dp\to\taup\nut with a statistical significance above \SI{4}{\stdDev}. 
The analysis of the $\Dz\to\KSL\piz(\piz)$ includes the measurement of the branching fractions and using the decays to $\KSL\piz$ the measurement of the \Dz mixing parameter $y_{\CP}$.
\section*{References}

\end{document}